\documentclass[conference]{IEEEtran}

\ifCLASSINFOpdf
   \usepackage[pdftex]{graphicx}
   \graphicspath{{../pdf/}{../jpeg/}}
   \DeclareGraphicsExtensions{.pdf,.jpeg,.png}
\else
   \usepackage[dvips]{graphicx}
   \usepackage{siunitx}
  
   \graphicspath{{../eps/}}
   \DeclareGraphicsExtensions{.png}
\fi

\usepackage{subfig}

\usepackage{array}
\usepackage{makecell}
\setcellgapes{4pt}
\usepackage{amsmath}\DeclareMathOperator*{\argmax}{argmax}
\usepackage{epstopdf,epsfig}
\usepackage{graphicx}
\usepackage{latexsym}
\usepackage{amsfonts}
\usepackage{epsfig}
\usepackage{subfig}
\usepackage{verbatim}
\usepackage{bibentry}
\usepackage{cite}
\usepackage{enumerate}
\usepackage{amsmath}    
\usepackage{floatflt}
\usepackage{float}
\usepackage{url}
\usepackage{lettrine}
\usepackage{array, makecell}

\makeatletter
\newcommand*{\rom}[1]{\expandafter\@slowromancap\romannumeral #1@}
\makeatother
\newcolumntype{M}[1]{>{\centering\arraybackslash}m{#1}}
\newcolumntype{P}[1]{>{\centering\arraybackslash}p{#1}}
\usepackage[utf8]{inputenc}

\usepackage{multirow}

\begin{document}
%
\title{ Pitch-synchronous Single Frequency Filtering Spectrogram for Speech Emotion Recognition }

\author{\IEEEauthorblockN{ Shruti Gupta  }
\IEEEauthorblockA{Department of Computer Science\\
National Institute of Technology\\
Patna, India \\
Email: shrgupta01@gmail.com}

\and
\IEEEauthorblockN{ Md. Shah Fahad}
\IEEEauthorblockA{Department of Computer Science \\
National Institute of Technology\\
 Patna, India\\
Email: shah.cse16@nitp.ac.in}

\and
\IEEEauthorblockN{Akshay Deepak}
\IEEEauthorblockA{Department of Computer Science \\
National Institute of Technology\\
 Patna, India\\
Email: akshayd@nitp.ac.in}

}


%

\maketitle

\begin{abstract}
Convolutional neural networks (CNN) are widely used for speech emotion recognition (SER). In such cases, the short time fourier transform (STFT) spectrogram is the most popular choice for representing speech, which is fed as input to the CNN. However, the uncertainty principles of the short-time Fourier transform prevent it from capturing time and frequency resolutions simultaneously. On the other hand, the recently proposed single frequency filtering (SFF) spectrogram promises to be a better alternative because it captures both time and frequency resolutions simultaneously. In this work, we explore the SFF spectrogram as an alternative representation of speech for SER. We have modified the SFF spectrogram by taking the average of the amplitudes of all the samples between two successive  glottal closure instants (GCI) locations. The duration between two successive GCI locations gives the pitch, motivating us to name the modified SFF spectrogram as pitch-synchronous SFF spectrogram. 
The GCI locations were detected using zero frequency filtering approach. The proposed  pitch-synchronous SFF spectrogram produced accuracy values of 63.95\% (unweighted) and 70.4\% (weighted) on the IEMOCAP dataset. These correspond to an improvement of +7.35\% (unweighted) and +4.3\% (weighted) over state-of-the-art result on the STFT sepctrogram using CNN.
Specially, the proposed method recognized 22.7\% of the happy emotion samples correctly, whereas this number was 0\% for  state-of-the-art results. These results also promise a much wider use of the proposed  pitch-synchronous SFF spectrogram for other speech-based applications.

\end{abstract}


%
\IEEEpeerreviewmaketitle

\section{Introduction}
 \lettrine[findent=0.5pt]{\textbf{S}}\sc peech emotion recognition (SER) refers to the classification/recognition of the person’s emotional state using the speech signal. 
 SER has a lot of applications in real life. It can be beneficial for applications where natural human-computer interaction is required. In computer tutorial applications, the detected emotion of the user can help the system in responding to the user's query \cite{el2011survey}. It can be incorporated in the onboard system of a car for initiating the safety of a passenger depending on his mental state \cite{schuller2004speech}. Medical professionals can use it as a diagnostic tool in psychological treatment \cite{france2000acoustical}. In automatic translation systems, it can help in effectively conveying the emotions between two parties\cite{akagi2014toward}. 
 
 For recognizing emotions from a speech, we need to extract emotion specific features, which are invariant with text\cite{wu2018text}. After the introduction of deep  neural networks, SER has gone through a significant growth over the past few years. There are two ways for extracting  features from speech: (i) hand-crafted features and (ii) deep neural network based features. Popular hand-crafted features are formant locations/bandwidths, pitch, voice probability,  zero-crossing rate, harmonics-to-noise ratio, mel filter bank features, mel frequency cepstral coefficients (MFCCs), energy, and jitter \cite{li2013automatic,meinedo2010age,shriberg2005modeling,ahmad2016gender}. There are two disadvantages with hand crafted features. First is the inherent inaccuracies  in extraction of these features. Second, the handcrafted features tend to overlook the higher level features that can be derived from the lower level features. On the other hand, in deep learning features are learned in a hierarchical manner; learning higher level abstractions  of the low-level features.

 Recently convolution neural networks (CNN) have become very popular for SER. CNN is a data-driven feature extractor in which filters are learned using data itself \cite{neumann2017attentive,mao2014learning,trigeorgis2016adieu,badshah2017speech,badshah2019deep}.  Short time  Fourier transform (STFT) spectrogram is widely used as an input to the CNN for speech applications \cite{mao2014learning, satt2017efficient, badshah2017speech, fayek2017evaluating }. Spectrogram is a time-frequency representation of a signal. There are various existing works in SER \cite{ mao2014learning, satt2017efficient, badshah2017speech, fayek2017evaluating } in which spectrogram was given as input to a CNN. In \cite{mao2014learning} the spectrogram was  used  as input to a CNN with sparse auto-encoder to find salient features for SER. In \cite{badshah2017speech} experiments were conducted using pre-trained AlexNet model and a freshly trained CNN model. The pre-trained AlexNet model was fine-tuned for  transfer learning. The experimental results showed that the proposed approach based on the freshly trained CNN model were better than the pre-trained AlexNet model. In \cite{fayek2017evaluating} experiments were conducted on several CNN and LSTM-RNN architectures. The CNN architectures achieved better performance as compared to the other LSTM-RNN architectures. 
 
 Reference \cite{satt2017efficient} highlights the drawback of spectrogram representation for noisy data. The authors modified the spectrogram using pitch information for robust SER. They also commented on Mel-scale spectrograms vs. linear-spaced spectrogram. Mel-scale spectrograms remove the pitch information but the emotions are strongly correlated with the pitch information. Therefore,  linear-spaced spectrogram is usually used for speech emotion recognition.  In \cite{Yenigalla2018SpeechER} spectrogram along with phoneme embedding features are combined with a CNN model to retain emotional contents of speech.  Both spectrogram and hand-crafted features have been used as input to the CNN for SER.  However, as noted in \cite{mirsamadi2017automatic}, the raw spectrogram has higher accuracy than the hand crafted features because the latter are already decorrelated.

 Speech signal is a non-stationary signal, i.e.,  it's behaviour changes with time. It is processed  frame-wise, where the size of a frame is small enough (typically around 25 ms) to assume that the characteristics of speech is stationary within a frame. In spectrogram, Fourier transform of a short-time windowed signal is taken with some overlapping (typically around 10 ms).  The shorter sized window (wide-band spectrogram) contains high temporal resolution but less spectral resolution, whereas the larger sized window (narrow-band spectrogram) contains high spectral resolution but less temporal resolution \cite{bayya2013spectro}. In other words, STFT spectrogram is unable to obtain high resolutions of time and frequency simultaneously. Since, the pattern of emotional speech varies rapidly within a glottal cycle due to the opening and closing of the glottis  \cite{kadiri2017epoch, yadav2018epoch},  it requires high temporal resolution for further analysis. 
 
  Wavelet transform \cite{kadambe1992application} is another time frequency representation that overcomes the limitation of the STFT spectrogram to some extent. Here, the time resolution is better in the high frequency region, while frequency resolution is better in the low frequency region.  However, in wavelet transform, the type of wavelets have to be chosen a priori, which often leads to misinterpretation of the data. On the other hand, the newly proposed single frequency filtering (SFF) spectrogram \cite{aneeja2015single,aneeja2017extraction,pannala2016robust,kadiri2017epoch} has both high temporal and spectral resolution without any requirement of a priori choices. Further, in absence of sensitive parameters like window-size (as in the case of STFT), SFF is more robust. SFF technique  has been successfully used in speech/non-speech detection \cite{aneeja2015single}, fundamental frequency extraction \cite{aneeja2017extraction, pannala2016robust} and glottal closure instants (GCIs) detection \cite{kadiri2017epoch}. To the best of our knowledge SFF has not been used for SER. 
  
 SFF spectrogram is especially appealing for SER because the behaviour of emotional speech is more non-stationary than normal speech. SFF output contains both high temporal and spectral resolutions. Hence, the SFF spectrogram captures transient parts of the emotional speech very well. The high spectral resolution clearly represents harmonic structures. The harmonics patterns, called timbre spectrum, are important for emotion classification \cite{klasmeyer1997perceptual,he2010importance}.   
  However, there is a practical problem in using SFF spectrogram for SER. SFF spectrogram extracts the amplitude envelope at each sampling instant. This leads to a huge feature matrix, making it computationally prohibitive to be used as the input to a CNN. Hence, a logical way is required to reduce the size of the feature matirx of SFF spectrogram. 

  Based on the above discussion, there are two primary motivations of our work: 
\begin{itemize}
\item \textbf {Use of SFF Spectrogram for SER}. The advantages offered by the SFF spectrogram over the STFT spectrogram and the wavelet transform technique make a strong case for exploring SFF spectrogram for SER.  
\item \textbf {Reduction of the size of the feature matrix of SFF.} 
A logical way is required to reduce the size of the feature matirx of SFF spectrogram to make its use computationally practical.  
\end{itemize}
  
  In light of the above motivations, the contributions of our work are as follows:
\begin{itemize}
\item \textbf {Pitch-synchronous SFF spectrogram:} We have proposed a novel pitch-synchronous SFF spectrogram, where the amplitude is averaged over pitch cycles. As compared to block processing -- where the assumption that the emotional speech is  stationary over fixed-size frames does not hold good practically -- the proposed approach logically decomposes frames based on GCI locations; an emotional speech is near stationary between two successive GCI locations. As a result, the unique pitch pattern of every emotion is reflected well in the pitch-synchronous SFF spectrogram (details in Sec. \ref{Sec:ResultDiscussion}).  To overcome the problem of a huge feature matrix, we  have averaged the amplitudes over pitch cycles, hence, reducing the size of the feature matrix. This also motivated us to name the modified SFF spectrogram as pitch-synchronous SFF spectrogram. 
\item \textbf {Improved performance:} We have compared pitch-synchronous SFF spectrogram with (i) STFT spectrogram and (ii) SFF spectrogram with 20 ms fixed sized frames (called ``SFF-20 ms spectrogram''). All representations were evaluated on a developed CNN model. The SFF-20 ms spectrogram achieves an improvement of 2.49\% for unweighted accuracy and 1.74\% for weighted accuracy over state-of-the-art STFT spectrogram using CNN \cite{satt2017efficient}. Pitch-synchronous SFF spectrogram further improves upon this by achieving improvements of 7.35\% for unweighted accuracy and 4.3\% for weighted accuracy over state-of-the-art STFT spectrogram using CNN \cite{satt2017efficient}.

\end{itemize}


 The rest of the paper is structured as follows. In Section \rom{2}, the proposed pitch-synchronous SFF spectrogram and CNN architecture are discussed in detail. Section \rom{3} describes experimental setup. Results are discussed in Sec. \rom{4}. Section \rom{5} concludes the paper with future directions.

\section{Proposed Approach} 


Our proposed work utilizes the SFF spectrogram \cite{aneeja2015single,aneeja2017extraction,pannala2016robust,kadiri2017epoch}, which is a time-frequency representation, for SER. The SFF spectrogram derives the amplitude envelope of a signal at each frequency as a function of time. In this way, the SFF spectrogram captures the temporal variation at each sample in an emotional speech. Note that this variation can also be captured by  the discrete Fourier transform (DFT) over a block of data at every sampling instant. But, this process is computationally expensive since the DFT is performed at each sample. The SFF spectrogram extracts the amplitude envelope at each sampling instant; resulting in a huge feature matrix. Computationally, it is impractical to use such a huge feature matrix as an input to CNNs. To overcome this, the amplitude envelope is averaged for all the samples between two successive GCI locations (called the pitch period). The resultant output representation is called pitch-synchronous SFF spectrogram. In section \ref{num1}, the detailed steps of constructing the pitch-synchronous SFF spectrogram are described. In Sec. \ref{num2} glottal closure instants (GCI) detection using zero frequency filtering (ZFF) method is described. In Sec. \ref{deepl} the proposed deep CNN architecture is described.

\subsection{Pitch synchronous SFF spectrogram} \label{num1}
Following are the steps to obtain the SFF spectrogram output.
\begin{enumerate}
\item  The speech signal $s[n]$, sampled at frequency $f_{s}$ Hz, is pre-emphasized ($p[n]$) to remove the low frequency bias from the speech signal.
\begin{equation}
p[n] = s[n]-s[n-1]
\end{equation}
\item  The pre-emphasized speech signal $p[n]$ is multiplied by a complex sinusoid of normalized shifted frequency as follows:
\begin{equation}
p[\bar{k},n] = p[n] e^{j\bar{\omega_{k}}n}
\end{equation}
where,\\
$p[\bar{k},n]$ is the resultant output at $n^{th}$ sample and \\ 
$\bar{\omega_{k}}$ is the normalized version of frequency $f_{k}$ at $k^{th}$ filter. $\bar{\omega_{k}}$ is computed as: 
  \begin{equation}
\bar{\omega_{k}} = \frac{2\pi \bar{f_{k}}}{f_{s}}
\end{equation}
where,\\
$\bar{f_{k}}$ is the shifted frequency shifted as $\bar{f_{k}} = f_{s}/2 -f_{k}$.

In the above, $n\in {1...N}$ and $k\in {1...K}$, where $N$ is the total number of samples and $K$ is the total number of filters. If the spacing between the filters is $\Delta f$ Hz, the total number of filters are $K = \frac{f_{s}/2}{\Delta f}$.

\item The resultant output $p[\bar{k},n]$ is passed through a single pole filter at $ f_{s}/2$ whose transfer function is given by
\begin{equation}
H(z) = \frac{1}{1+rz^{-1}}
\end{equation}
The stability of the filter is ensured by choosing the value of the pole of the filter inside the unit  circle (radius $r=0.9394)$.  The output of SFF gives high resolution at each frequency because it considers the effect of one resonator at a time and reduces the effect of other resonators significantly.
\item  The output of the filter is given by
\begin{equation}
y[k,n] = -ry[k,n-1]+p[\bar{k},n]
\end{equation}
$y[k,n]$ is a complex number with real part $y_{r}[k,n]$ and imaginary part $y_{i}[k,n]$.
\item The SFF envelope, denoted $e[ k, n ]$, of the filtered output $y[ k, n ]$ at the $k^{th}$ filter is given by
\begin{equation}
e[k,n] = \sqrt{y^{2}_{r}[k,n]+y^{2}_{i}[k,n]}
\end{equation}
In out work, the SFF envelope is calculated for frequency range $[0, f_{s}/2]$ Hz with $20$ Hz spacing. The SFF envelope of an anger utterance is shown in Fig \ref{seven}.

\item The GCI locations from the speech signal $s[n]$ are detected using the ZFF algorithm \cite{murty2008epoch} described in Sec. \ref{num2}. Let $z[n]$ denote the ZFF signal and  $s[l]$ denote the set of GCI locations extracted from   $z[n]$.

\item Subsampling: It refers to the process of reducing the number of samples  by  averaging the samples  between two  successive GCI locations in the SFF envelope. The average values are computed as: 
\begin{equation}
u[k,l] =\frac{1}{s[l+1] - s[l]}\sum_{i=s[l]}^{s[l+1]} e[k,i] ,  \forall s[l]
\end{equation}
 
where $s[l]$ is the GCI location and $i$ iterates over the samples between successive GCI locations $s[l]$ and $s[{l+1}]$.  The resultant output is called pitch-synchronous SFF envelope.
\begin{figure}
\centering
\includegraphics[width=8.5 cm]{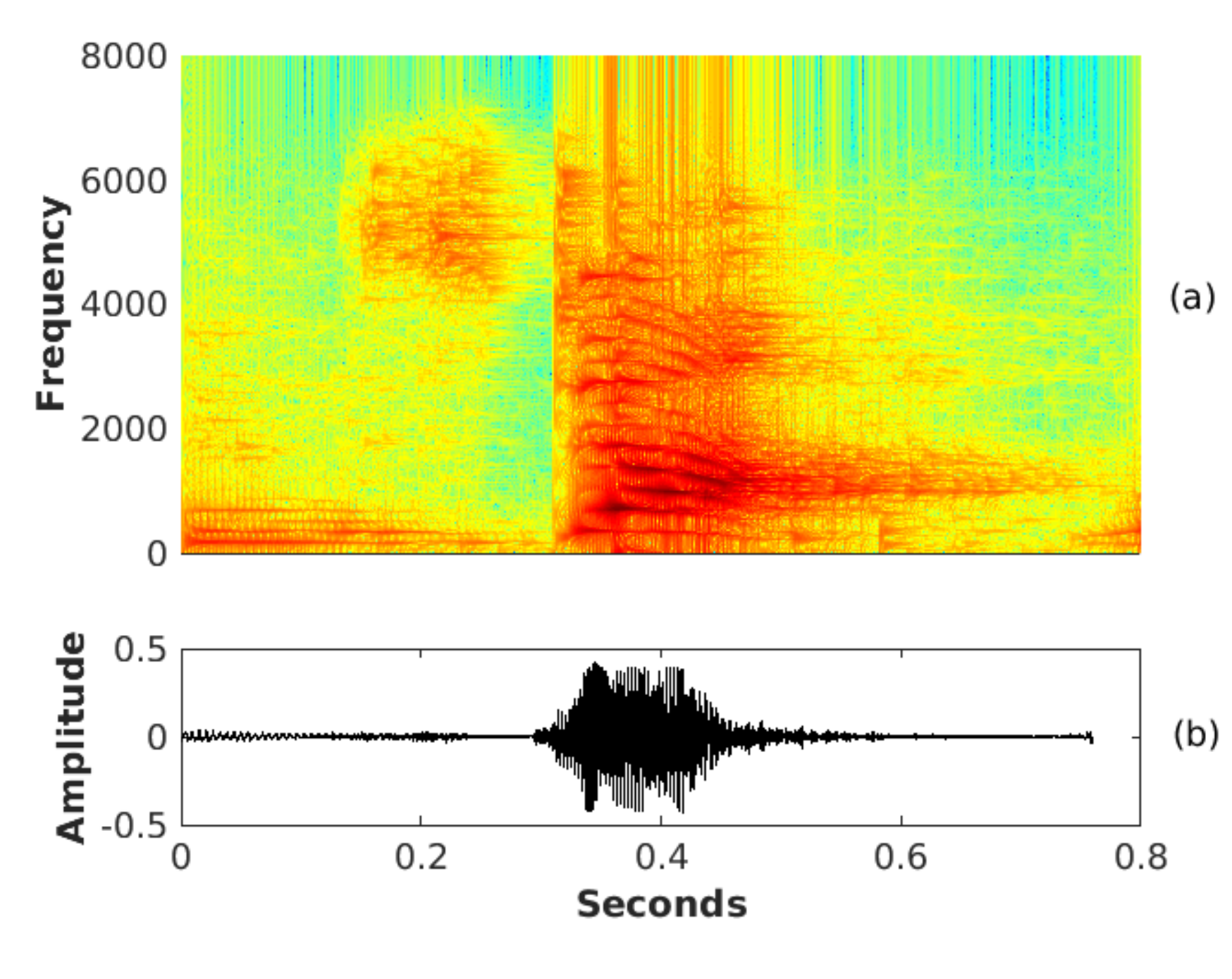}
\caption{SFF time-frequency representation. The SFF spectrogram of an angry utterance (b) is shown in (a) }\label{seven}
\end{figure}

\item Finally, the log-spectrum of the pitch-synchronous SFF envelope is computed as follows:
\begin{equation}
x[k,n] = log(u[k,l])
\end{equation}
\end{enumerate}
 
\begin{figure*}
 \begin{center}
\centering
 \includegraphics[width=14 cm]{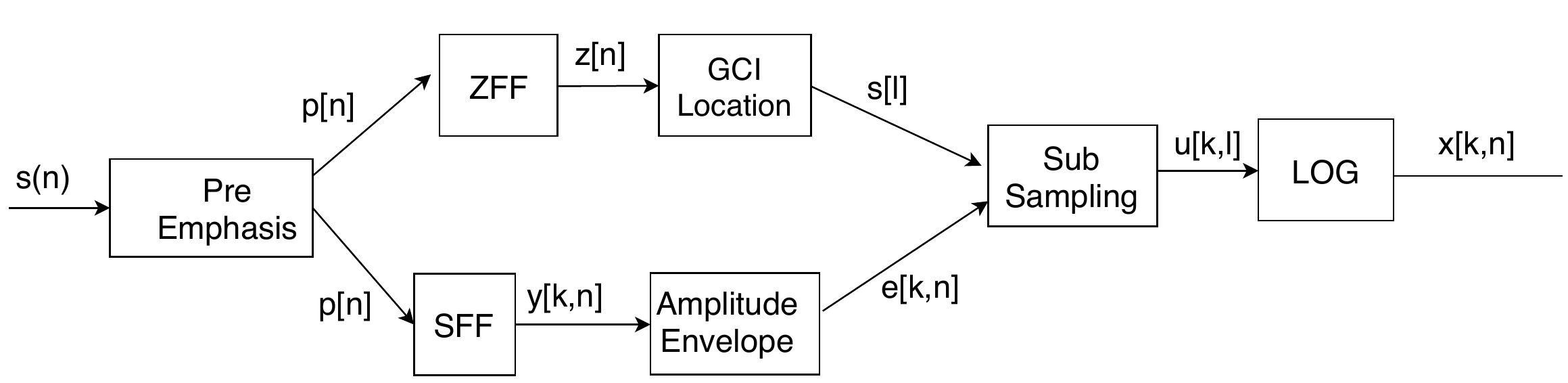}
\caption{The block diagram of pitch-synchronous SFF method. $s[n]$ is the speech signal, which is further pre-emphasized. The pre-emphasized signal $p[n]$ is given to both the ZFF and SFF blocks. The output of the ZFF block is the ZFF signal $z[n]$, from which the GCI locations $s[l]$ are extracted. The output of the SFF block is $y[k,n]$, from which the amplitude envelope $e[k,n]$ is derived. The SFF envelope is subsampled by averaging the samples between two consecutive GCI locations. The log of the resultant pitch-synchronous SFF $u[k,l]$ gives the final output $log(u[k,l])$. }\label{six}
\end{center}
\end{figure*}

The flow diagram of the proposed pitch-synchronous SFF spectrogram is shown in Fig \ref{six}.
   


\subsection{ GCI detection using the Zero Frequency Filtering (ZFF) method}  \label{num2}
The GCI locations are detected by the ZFF method proposed in
 \cite{murty2008epoch}. Following are the steps to detect the GCIs using the ZFF method.
 \begin{enumerate}
 \item   Speech signal is differentiated to pre-emphasize the high frequency components as:
   \begin{equation}
   p[n] = s[n]-s[n-1]
   \end{equation}
The resultant signal is denoted as $p[n]$ and known as pre-emphasized signal. 
   \item The pre-emphasized signal $p[n]$ is passed through a zero frequency resonator. The response of the zero frequency resonator is defined as:
   \begin{equation}
    z_{0}[n] = -\sum_{i=1}^{2}a_{i} z_{0}[n-i]+p[n]
   \end{equation}
   where $ z_{0}[n]$ is the resonator output, $a_{1} = -2$ and $a_{2}=1$.
  
   \item  For removing the trend in the resonator output $z_{0}[n]$ a moving average filter is used whose window length corresponds to the pitch period of that utterance. The resulting output is called the ZFF signal:
   \begin{equation}
    z[n] = z_{0}[n] -\sum_{m=-M}^{M} z_{0}[n+m]
   \end{equation}
   where $2 M + 1$ corresponds to the number of samples used for computing the trend.
   \item  The positive zero crossing of the ZFF signal $z[n]$ corresponds to the epoch location and is denoted by $s[l]$.
 \end{enumerate}



\begin{figure}[h]
\centering
\includegraphics[width=8.3 cm,height = 8cm]{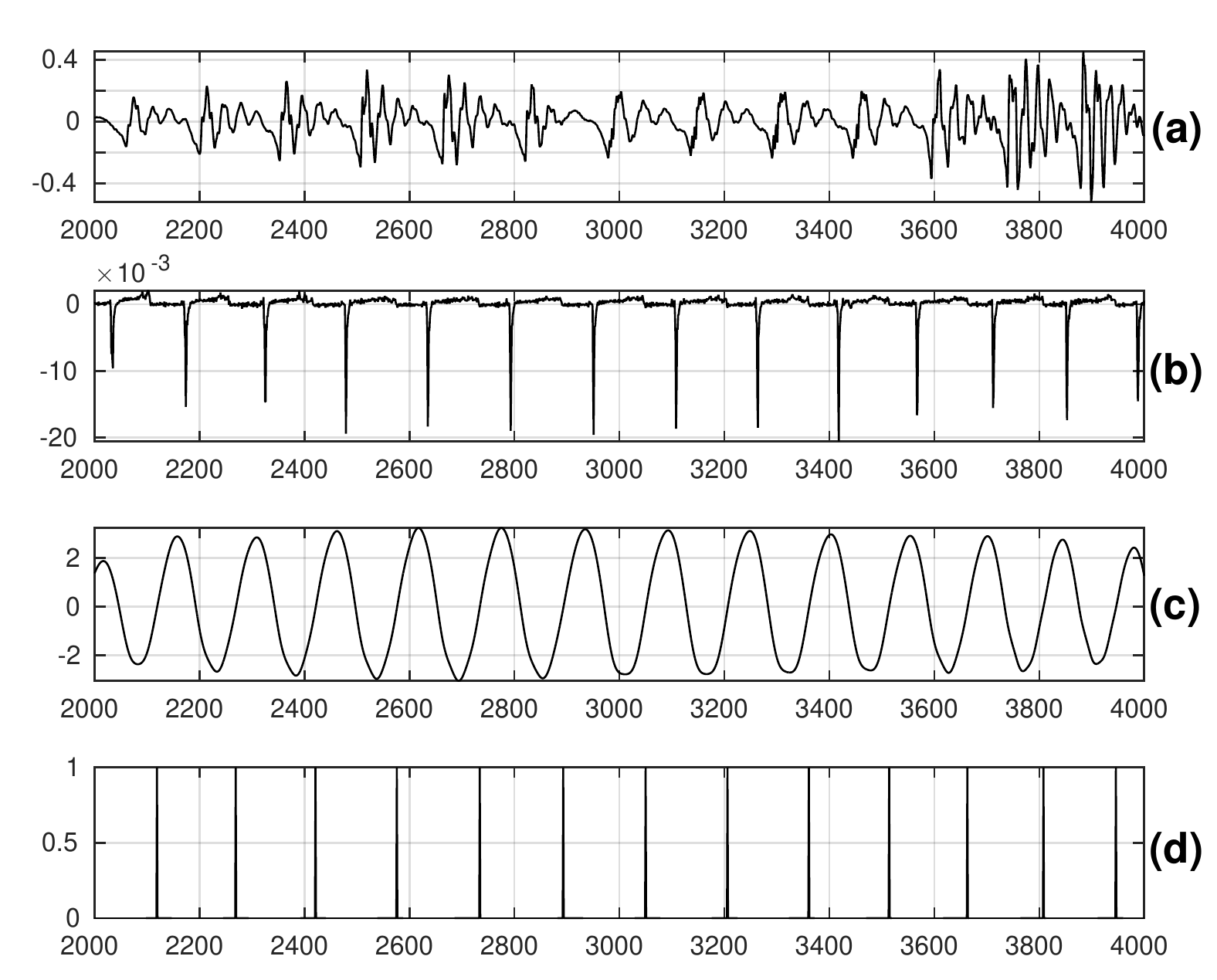}
\caption{GCI detection using ZFF method. (a) is the voiced segment (b) is the corresponding DEGG signal. (c) Zero frequency filtered signal. (d) Epoch location.}\label{one}
\end{figure}

Figure 3(a) shows the voiced segment of a sentence . Fig. 3(b) shows the differentiated eloctro-glottograph (DEGG) signal of the voiced segment of 3(a).  Figure 3(c) shows the zero frequency filtered (ZFF) signal and  Fig. 3(d) shows the positive crossing of the ZFF signal which corresponds to the epoch location.

\subsection{Deep CNN Architecture}  \label{deepl}
Deep learning helps in learning multiple levels of abstraction by composing networks into multiple layers. The network learns features in a hierarchical manner. Low-level features are learned from the raw data and higher level features are learned from the lower level features. Thus, deep learning overcomes the dependency of shallow networks on hand-crafted features by learning the features itself. In general, a convolutional neural network (CNN) is composed of convolution layers and pooling layers. A convolution layer learns the  filters for low-level features by itself, whereas traditionally, these filters were designed by human experts. Due to this advantage of automatically learning the features, raw inputs are directly fed to a CNN and it extracts the features from the input. The max-pooling  layers extract dominant features that are robust against distortion and noise. Additional convolution and max-pooling layers are capable of deriving more complex features from intermediate features. A CNN is composed of repetitive units called CNN blocks. In this section the architecture of the CNN block used in our work and the learning procedures are discussed.

\subsubsection{CNN block}
\begin{figure}[h]
\centering
\includegraphics[width = 3.5cm,angle =90]{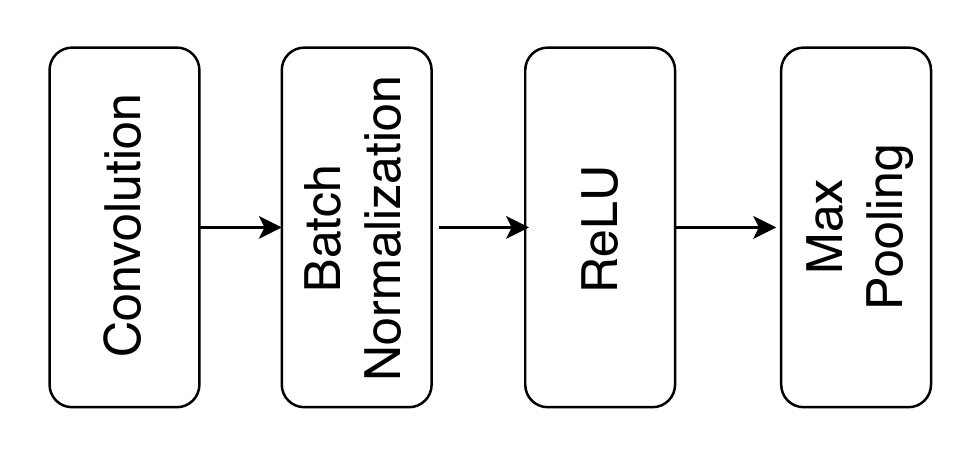}
\caption{CNN Block}\label{three}
\end{figure}

  The CNN block used in our work consists of four components: one convolution layer, one batch normalization (BN) layer \cite{ioffe2015batch}, one rectified exponential linear unit (ReLU) layer \cite{nair2010rectified}, and one max-pooling layer, in that order, as shown in Fig. \ref{three}. The core layers of the CNN block are convolution layer and pooling layer. The BN layer helps in improving the stability and performance of deep neural networks. It normalizes the activations at each batch and maintains zero mean and unit variance \cite{ioffe2015batch}. ReLU layer uses ReLU activation function. It is widely used because it can lead to higher recognition accuracies and faster convergence rates \cite{nair2010rectified}. Max-pooling layer uses a max-filter that is applied to the sub-regions of the output of the convolution layer  \cite{cirecsan2012multi}. Depending upon the requirements, these CNN blocks can be configured accordingly.

In a convolution layer, each index value $s(m,n)$ of the input $s$, which is a 2D matrix, is convolved with the convolution kernels. The input for the first layer is the raw spectrogram. The input for the convolution layers in the subsequent CNN blocks is the output of the previous CNN block.  The convolution operation is performed between the convolution kernel and the input data to produce the feature maps.  The convolution kernel $w(m,n)$ of size $(x, y)$ is initialized randomly. When  $s(m, n)$ is convolved with convolution kernel $w(m, n)$, the resultant output $O(m, n)$ is obtained as:
\begin{equation}
\begin{aligned}
O(m, n) &= s(m,n) * w(m,n) \\
        &= \sum_{a=-x}^{a=+x} \sum_{b=-y}^{b=+y} s(a, b) \cdot w(m-a,n-b) \\      
\end{aligned}
\end{equation}

The resultant output is given as input to the BN layer. The features that are learned by a convolution layer are normalized by the corresponding BN layer. The output of a BN layer is given to the corresponding ReLU layer as follows: 
\begin{equation}
O_m^l = \sigma(BN(b^l_m + \sum_n O_n^{l-1} * w_{mn}^l ))
\end{equation}
where, \\
$O_m^l$ represents the $m^{th}$ output feature at the $l^{th}$ layer, \\
$O_n^{l-1}$ represents the $n^{th}$ input feature at the $(l-1)^{th}$ layer, and \\
$w_{mn}^l$ denotes the convolution kernel between the $m^{th}$ and $n^{th}$ feature.

The ReLU activation function $\sigma$(.) of the network is defined as follows:

\begin{equation}
 \sigma(x) = \begin{cases} x & x \geq 0 \\
 0 & x < 0 \end{cases}
\end{equation}
where $x$ denotes the activation potential.

The next layer is max-pooling or downsampling layer. The normalized features are passed to this layer. The resolution of the features is reduced by taking the maximum value of a sub-region in the pooling layer. The output of a max-pooling layer can be defined as:
\begin{equation}
O^l = \max_{\forall p \epsilon \Omega} O_p^l
\end{equation}
where, \\
$O^l$ represents the output features,\\
$O_p^l$ represents the input features, and\\
$\Omega$ depicts the pooling patch.

\subsubsection{Learning Procedures}
 Learning of the neural network is phrased as an optimization problem to minimize the loss between the targeted and the predicted output. Our work is formulated as a multi-class classification problem. Therefore, the categorical cross-entropy loss\cite{de2005tutorial} function is used. The cross-entropy loss gives a probability value between zero and one to measure the performance of the corresponding classification model. As the predicted probability diverges from the target value, the cross-entropy loss increases. A separate loss for each class label per observation $i$ is calculated. All these losses are summed and the combined loss is obtained as follows:
\begin{equation}
L_i = - \sum_{x=1}^{C} i_{x,c} * log(p_{x,c})
\end{equation}

where, \\
 $C$ is the number of classes, \\
 $i$ is a binary indicator; it is one for correct classification and zero otherwise, and \\ 
 $p_{x,c}$ is the predicted probability of sample $x$ for class $c$. 
 
 The cross-entropy loss over a mini-batch is computed by taking the average:
\begin{equation}
L = \frac{1}{N} \sum^{N}_{i=1} L_{i}
\end{equation}
where $N$ is the number of samples in a batch.

For computing the gradients, the cost function is differentiated with respect to the model parameters and backpropagated to prior layers using the backpropagation algorithm \cite{leonard1990improvement,riedmiller1993direct}. 
After obtaining the gradients, in general, the gradient descent algorithm or one of it's variants is used for updating the model parameters.
 In our work, Adam optimizer \cite{kingma2014adam,reddi2018convergence} has been used for loss optimization.
Adam stands for adaptive moment estimation. Using Adam, for each parameter, adaptive learning rates are computed. In addition to computing an exponentially decaying weighted average of past gradients $m_t$, similar to momentum, Adam also computes an exponentially decaying weighted average of the squared gradients $v_t$, where $\beta_{1}$ and $\beta_{2}$ control the decaying rates of these moving averages as shown below: 
   \begin{equation}
   m_t = \beta_{1} * m_{t-1} + (1 - \beta_{1}) g_t
   \end{equation}
   \begin{equation}
   v_t = \beta_{2} * v_{t-1} + (1 - \beta_{2}) {g_t}^2
   \end{equation}
  
where $g_t$ denotes the gradient value at step $t$. The initial value of the moving averages  $\beta_{1}$ and $\beta_{2}$ is close to one. This results in a bias of moment estimates towards zero. To counteract this, a bias correction is applied as follows:

\begin{equation}
\hat{m_t} = \frac {m_t} {1 - \beta_{1}^{t}}
\end{equation}

\begin{equation}
\hat{v_t} = \frac {v_t} {1 - \beta_{2}^{t}}
\end{equation}
where, \\
$m_t$ is the first momentum representing the mean of gradients and\\
$ v_t$ is the second momentum representing the uncentered variance of gradients. 

 For updating the parameters, the final formula is given below:
\begin{equation}
\theta_{t+1} = \theta_{t} - \frac {\eta} {\sqrt{\hat{v_t}} + \epsilon}\hat{m_t}
\end{equation}
where, \\
$\theta$ is the parameter and\\
$\eta$ is the learning rate.

The classifier used in this architecture is softmax. Softmax is a generalization of logistic regression for multi-class classification. The softmax classifier gives the probabilities for each class label. The softmax function is defined as:
\begin{equation}
z_i = \sum_j h_j W_{ji}
\end{equation}

\begin{equation}
\sigma(z)_i = \frac{e^{z_i}} {\sum_{j=1}^{n} e^{z_j}}
\end{equation}
where, \\
$z_i$ is the input to softmax, \\
$W_{ji}$ is the weight between the
penultimate layer and softmax layer, and \\
$h_j$ is the activation in the
penultimate layer. 

The class 
label of each segmented utterance  $\hat{y_t}$ is predicted as follows:

\begin{equation}
\hat{y_t} = \argmax_{i} p_i
\end{equation}
where $p_i$ is the probability for class $i$.

\section{Experimental Setup}
Our proposed model has been evaluated on the IEMOCAP (Interactive emotional dyadic motion capture)\cite{busso2008iemocap} dataset. The IEMOCAP dataset is a multi-modal dataset which contains audio, video, text, and gesture information of conversations arranged in dyadic sessions. In this work, we have considered only the audio tracks of the IEMOCAP dataset. The IEMOCAP dataset is a natural-like improvised dataset that reasonably resembles spontaneous real life emotional speeches. 

The IEMOCAP dataset comprises of five sessions. In each session, there are two different speakers. There is no overlapping between speakers of different sessions. In each session, there is one male and one female. The conversation of one session is approximately five minutes long. The contents of the dataset are recorded in both scripted and spontaneous scenarios. In our study, we have considered four categorical (class) labelled emotions namely angry, happy, sad and neutral of improvised sessions\cite{busso2008iemocap}. We have used leave-one-speaker-out (LOSO) cross validation strategy. In LOSO  four sessions are used for training. One speaker form the remaining fifth session is used for validation while the the other speaker is used for testing.  This step is repeated for all the five sessions. 

In this dataset, all utterances are of different lengths; varying from one to twelve seconds approximately. To deal with the variance in the lengths of the input utterances,  the utterances are split into three second segments. Except the last segment, all segmented utterances are $T=3$ seconds long. Each segment is assigned the label of its parent utterance.  These segmented utterances are used for training and validation. During testing, to retrieve the label at the utterance level, the posterior probabilities of all the constituent segments is averaged. The class label with the highest average value is assigned as the corresponding utterance label:
      \begin{equation}
       s = \argmax_{n = 1...N} \frac {\sum_{t = 1}^{T} p(y_{t}|x)}{T}
      \end{equation}
      
where, \\
$s$ is the predicted label of the utterance,\\
$p(y_{t}|x)$ is the posterior probability of segment $t$,\\
$T$ is the number of segments, and \\
$N$ is the number of classes, i.e., four.

The pitch-synchronous SFF spectrograms are computed as described in Sec. \ref{num1}.  There are only two parameters that need to be set: the value of the radius of the filter $r = 0.9394$ and the spacing frequencies $\Delta f = 20 Hz$. The range of selected frequencies is $0-4 kHz$, which covers human speech. The GCI locations are identified as described in Sec. \ref{num2}. The maximum number of GCI locations found in our experiments in a $T=3$ second segment is $1077$. Therefore, the size of feature matrix for a segment is  $(200 \times 1077)$, where $200$ is the number of frequency bins. The feature matrix of segments having less than $1077$ GCI locations is padded with zeros to reach the fixed size of $(200 \times 1077)$. We compared the pitch-synchronous SFF spectrogram with the STFT spectrogram and the SFF- $20$ ms  spectrogram. In the STFT spectrogram, the speech signal is segmented into frames of $40$ ms with an overlapping of $10$ ms. Hamming window is applied on each frame to compensate the end effect of the frame. The discrete Fourier transform of each frame is calculated with the window length of 800 (for $20$ Hz). In SFF-$20$ ms, the averaging of the amplitude of all the samples is performed with $50$\% overlapping.


We used a convolutional neural network (CNN) for evaluating our proposed pitch-synchronous SFF. To ensure exactly the same experimental conditions, we have implemented the same CNN architecture as proposed in state-of-the-art method \cite{satt2017efficient}, except that we have used batch normalization for each convolution layer. The detailed parameters of each CNN block is given in Table \ref{T9}.
The used $2D$ CNN network has the following structure. It has three CNN blocks (CNN Block1, CNN Block2, and CNN Block3), one fully connected layer and the output layer with the softmax classifier. The convolution and pooling kernels in each CNN block are $2-D$. In CNN Block1, there are $16$ convolution kernels of size $12 \times 16$. CNN Block2 consists of $24$ kernels of $8 \times 12$ size and CNN Block3 has $32$ kernels of size $5 \times 7$. The
 sizes of the max-pooling layer  of the first, second and third CNN blocks is $100 \times 150$,  $50 \times 75$  and  $25 \times 37$ respectively.

The output obtained from the convolution layers  represents high-level features. The output of the last CNN block is passed through the fully connected layer to learn non-linear combinations of these features for SER.
The pitch-synchronous SFF spectrogram and the STFT spectrogram were implemented in MATLAB. The model is implemented using  Keras, which is a library widely used for deep learning.

\begin{table}[h]
\centering
\caption{Parameters of each CNN-Block}\label{T9}
\begin{tabular}{|l|c|c|c|}
\hline
\textbf{ Name} & \textbf{Layers} & \textbf{Kernel Size}  & \textbf{Output Size} \\
 \hline
 \multirow{2}{*}
 {CNN Block1} & Convolution2D & 12 $\times$ 16 & 189 $\times$ 284 $\times$ 16 \\\cline{2-4}
 & Max Pooling2D & 100 $\times$ 150 &  90 $\times$ 135 $\times$ 16\\

 \hline 
 \multirow{2}{*}
  {CNN Block2} & Convolution2D & 8 $\times$ 12 &  83 $\times$ 124 $\times$ 24 \\\cline{2-4}
 & Max Pooling2D & 50 $\times$ 75 &  34 $\times$ 50 $\times$ 24 \\
 
 \hline 
 \multirow{2}{*}
 {CNN Block3} & Convolution2D & 5 $\times$ 7 &  30 $\times$ 44 $\times$ 32 \\\cline{2-4}
  & Max Pooling2D & 25 $\times$ 37 & 6 $\times$ 8 $\times$ 32\\
 
 \hline
 \makecell{ Fully Connected \\Layer} &  & 64 &  64 \\

 \hline
 Output Layer & & 4 & 4 \\
 \hline
 
\end{tabular}


\end{table}

\section{Result Discussion}\label{Sec:ResultDiscussion}
The IEMOCAP dataset used in this work is significantly imbalanced.  To cope with the imbalance issue, weights are assigned to each class during training such that the weight of a class is inversely proportional to the total number of samples in that particular class.  The testing data is also imbalanced. Hence, we calculate unweighted as well as weighted accuracy. Unweighted accuracy is important for imbalanced data because it gives equal weightage to each class. Specifically, it is not biased against minority classes. Formally:
\begin{itemize}
\item
\textit{Overall accuracy} or \textit{weighted accuracy (WA)}  is defined as the number of samples predicted correctly out of the total number of samples.
\item
\textit{Average class accuracy} or \textit{unweighted accuracy (UWA)}  is defined as the average accuracy of individual classes. The individual class accuracy  is defined as the number of samples predicted correctly out of the total number of samples in that class. 
\end{itemize}

\begin{figure*}[htpb]
 \centering 
 \captionsetup[subfigure]{labelformat=empty}
 		\subfloat[Pitch-synchronous SFF]{\includegraphics[scale=0.495]{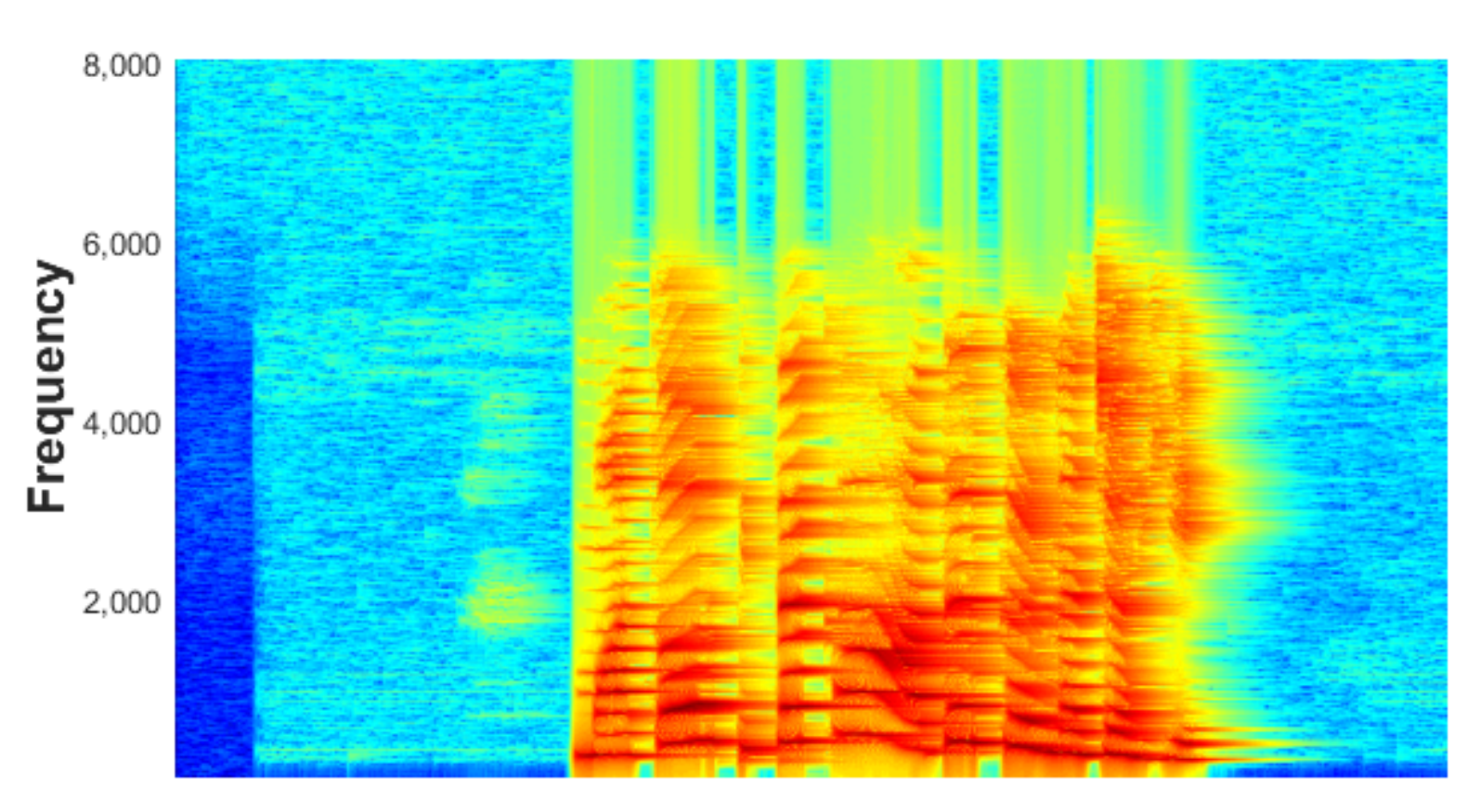}}
  	\subfloat[Pitch-synchronous SFF]{\includegraphics[scale=0.495]{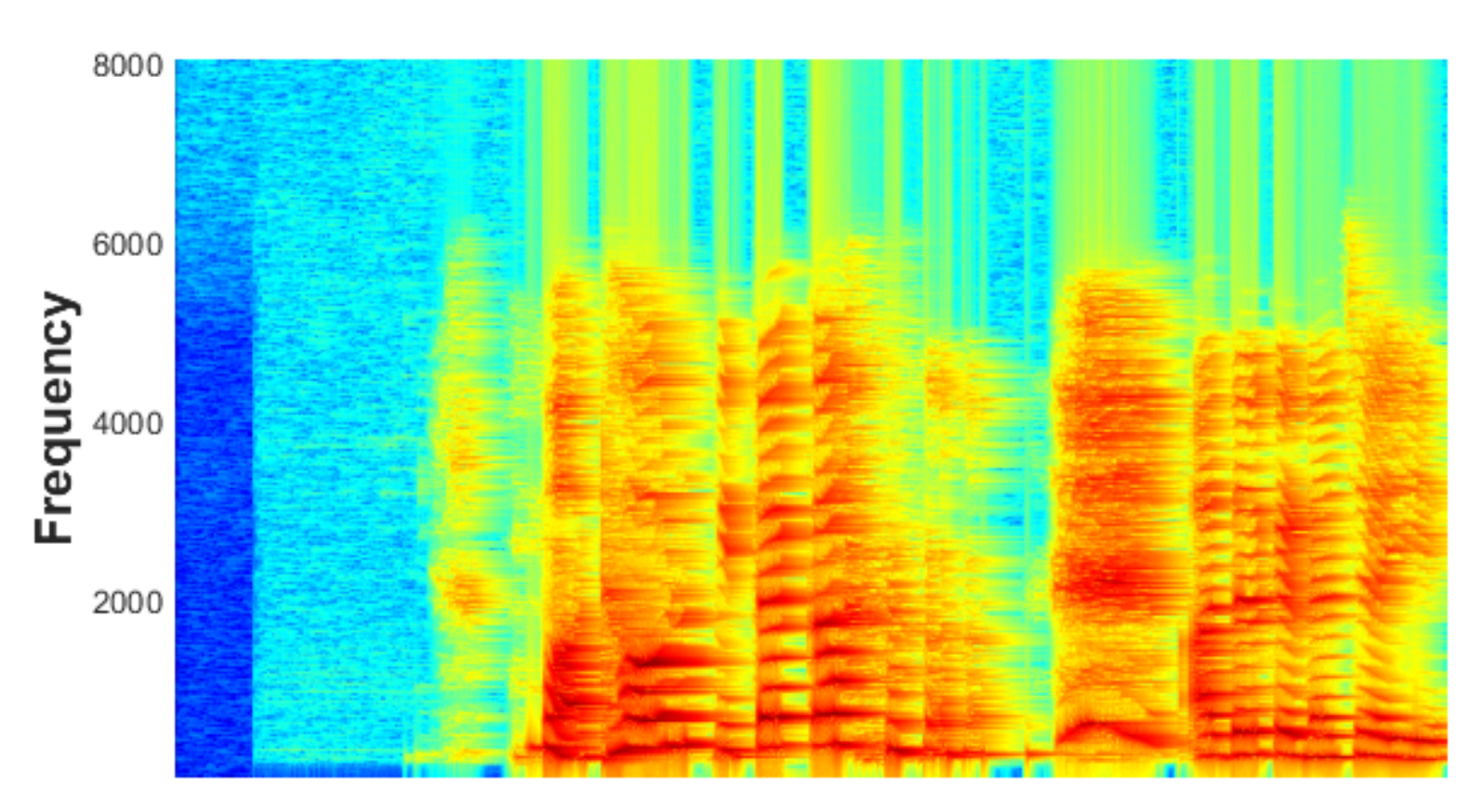}}\\
   \subfloat[$\>$ $\>$ $\>$ $\>$ $\>$ $\>$ $\>$ $\>$ $\>$ $\>$ $\>$ STFT Spectrogram \newline \newline (a) Anger]{\includegraphics[scale=0.495]{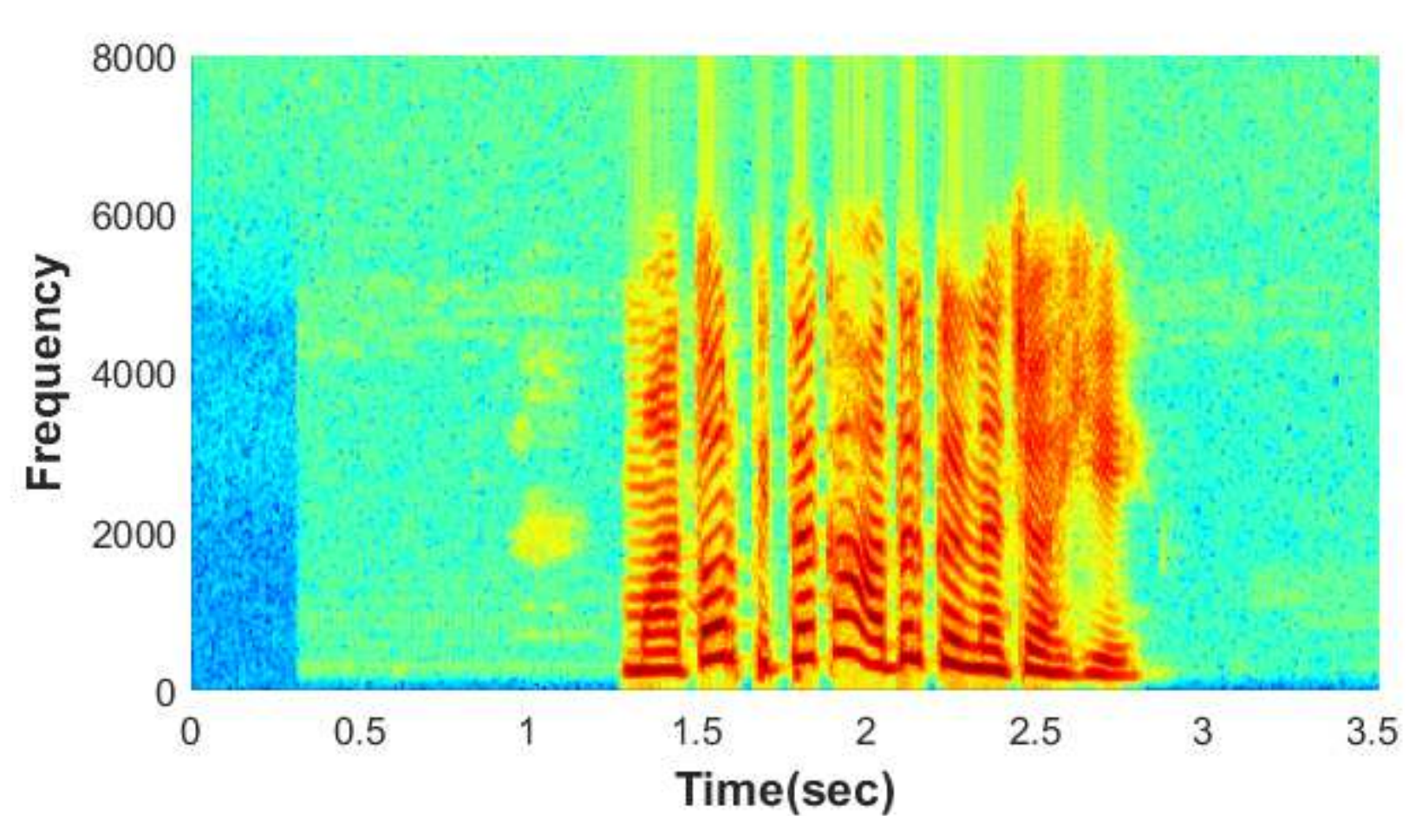}}
  \subfloat[$\>$ $\>$ $\>$ $\>$ $\>$ $\>$ $\>$ $\>$ $\>$ $\>$ $\>$ STFT Spectrogram \newline \newline (b) Happy]{\includegraphics[scale=0.495]{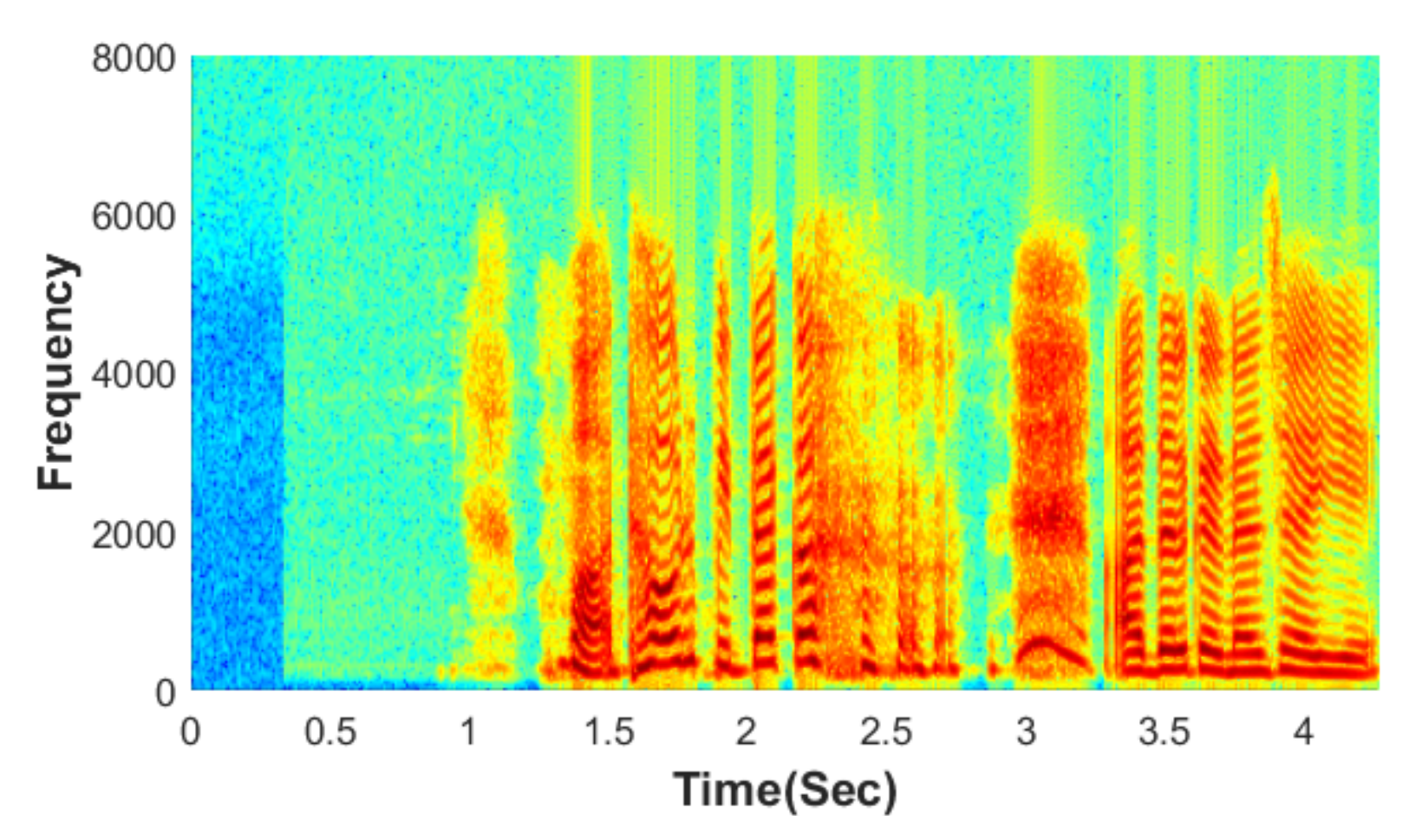}}\\
    

  
  \subfloat[Pitch-synchronous SFF]{\includegraphics[scale=0.495]{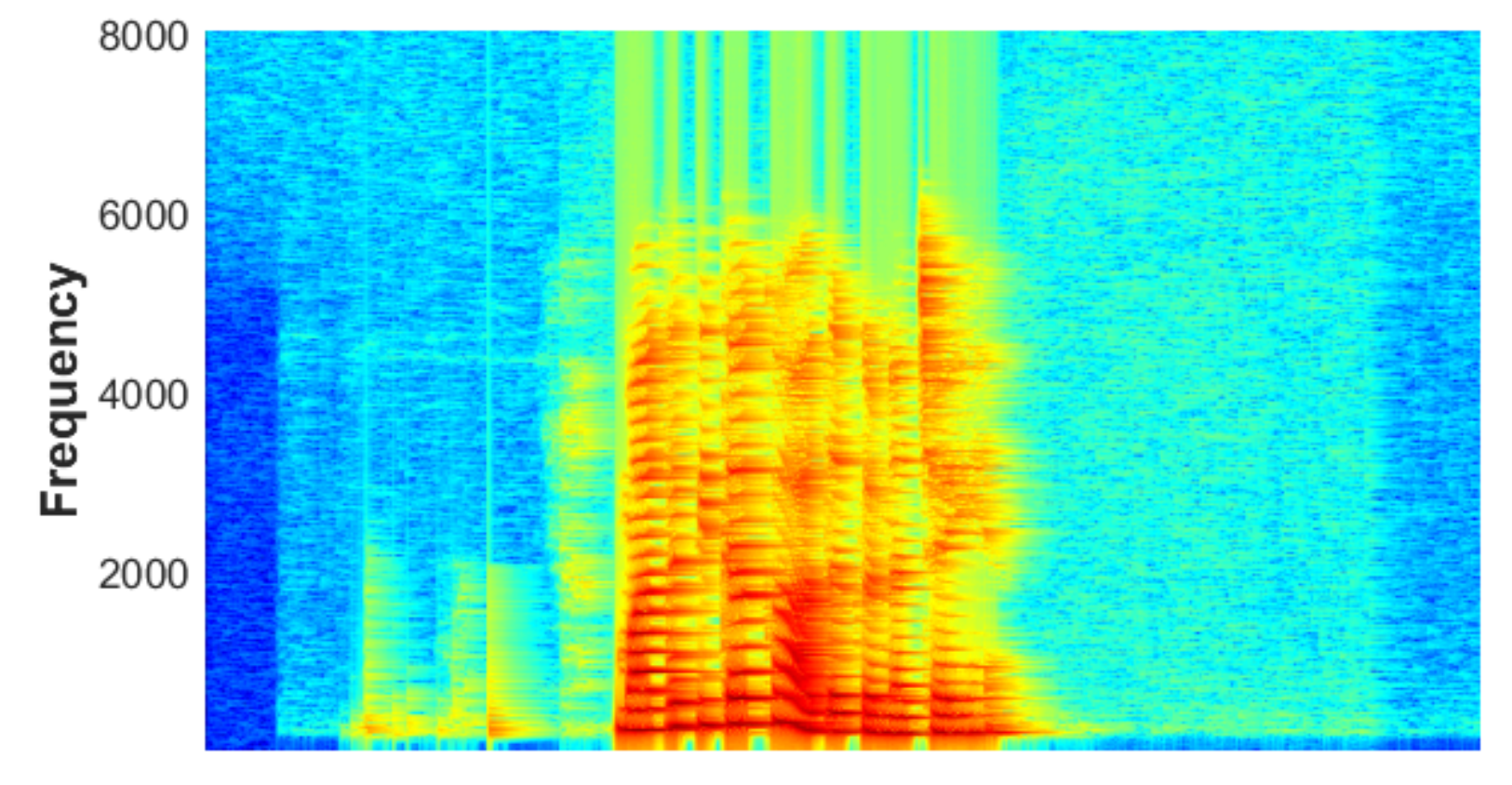}}
  \subfloat[Pitch-synchronous SFF]{\includegraphics[scale=0.495]{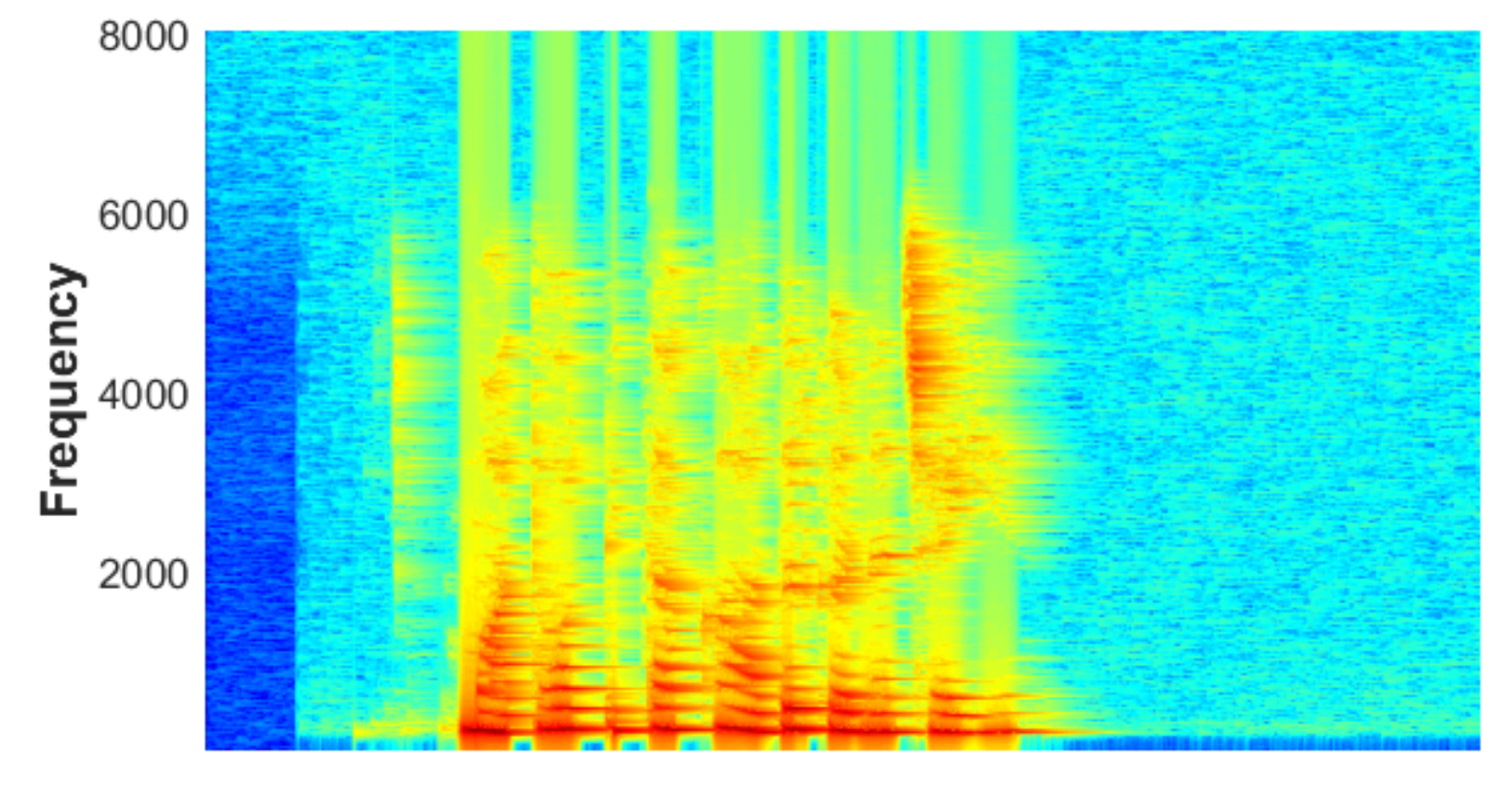}}\\
  \subfloat[ $\>$ $\>$ $\>$ $\>$ $\>$ $\>$ $\>$ $\>$ $\>$ $\>$ $\>$ STFT Spectrogram \newline \newline (c) Neutral]{\includegraphics[scale=0.495]{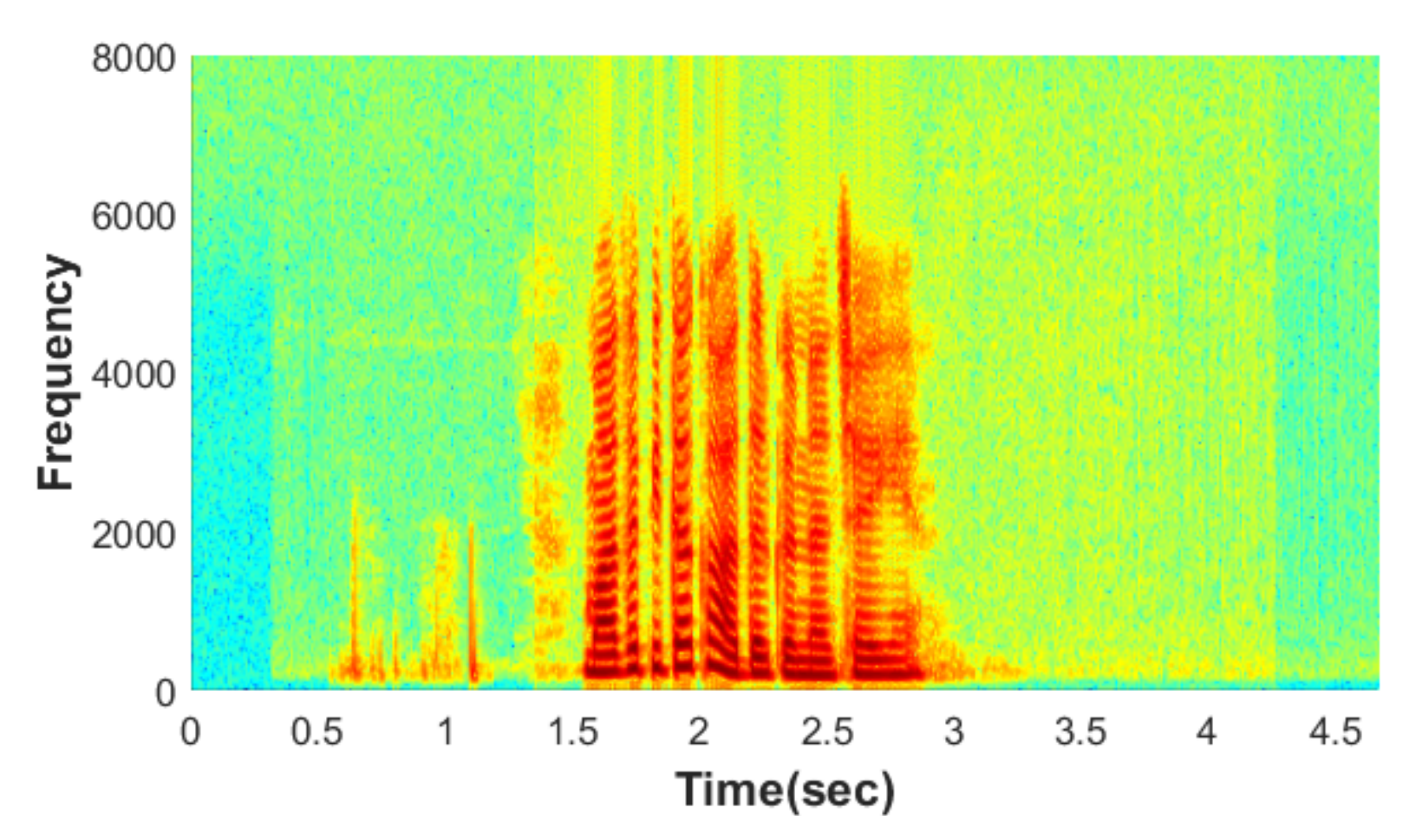}}
    \subfloat[ $\>$ $\>$ $\>$ $\>$ $\>$ $\>$ $\>$ $\>$ $\>$ $\>$ $\>$ STFT Spectrogram \newline \newline (d) Sad]{\includegraphics[scale=0.495]{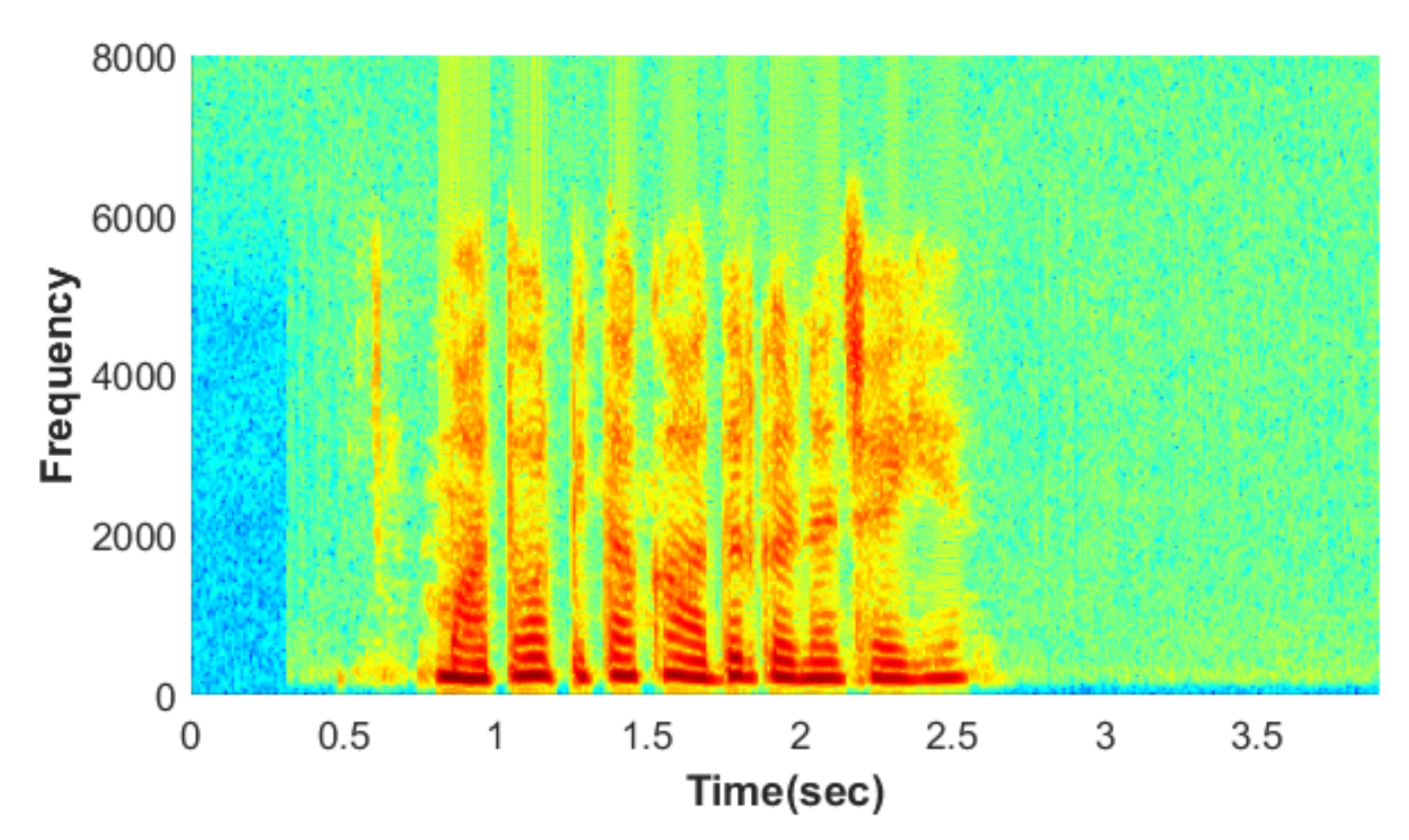}}\\
        \caption{The pitch-synchronous SFF and STFT spectrograms of the (a) anger, (b) happy, (c) neutral, and (d) sad emotions. In each of the sub figures, the top panel shows the corresponding pitch-synchronous SFF spectrogram while the bottom pannel shows the corresponding STFT spectrogram. }\label{five}
\end{figure*}

The comparison of the pitch-synchronous SFF spectrogram with the STFT spectrogram for the four considered emotions is shown in Fig. \ref{five}. All the spectrograms are drawn from the same sentence of the same speaker. Pitch-synchronous SFF is an adaptive framing approach in which the frame-size is derived from the instantaneous pitch-period in the corresponding utterance. It can be seen that the pitch-synchronous SFF spectrogram captures clearer harmonics than the STFT spectrogram. The pitch-synchronous SFF spectrogram also contains more fine grained temporal information than the STFT spectrogram. These properties of the pitch-synchronous SFF spectrogram enable better discrimination between anger and happy emotions, which are known to produce very similar characteristics with respect to the existing features. 
The pitch-synchronous SFF spectrograms of happy and anger emotions exhibit clear energy and formant structures than the neutral emotion. Figure \ref{five} (c) \& (d)  show the STFT spectrogram and the pitch-synchronous SFF spectrogram of neutral and sad emotions. The STFT spectrogram produces similar pattern for  neutral and happy emotions, i.e., the energy concentrations are same across the entire frequency region. This is not good for discriminating among these two emotions. However, the pitch patterns of happy emotion are prominently different than the neutral emotion in the case of the pitch-synchronous SFF spectrograms. This results in a superior discrimination of happy and neutral emotions.

 In the last layers of CNN architecture, the relationship between the feature maps are abstracted and the meaningful timbre spectrum is obtained. The GCI locations are present in the voiced regions. A frame is constructed based on the GCI locations, which automatically reduces the effect of silence region. We can clearly see in all the pitch-synchronous SFF spectrograms that the effect of silence between words is highly reduced.

\begin{figure*}[t]
\centering

\includegraphics[width = 15cm]{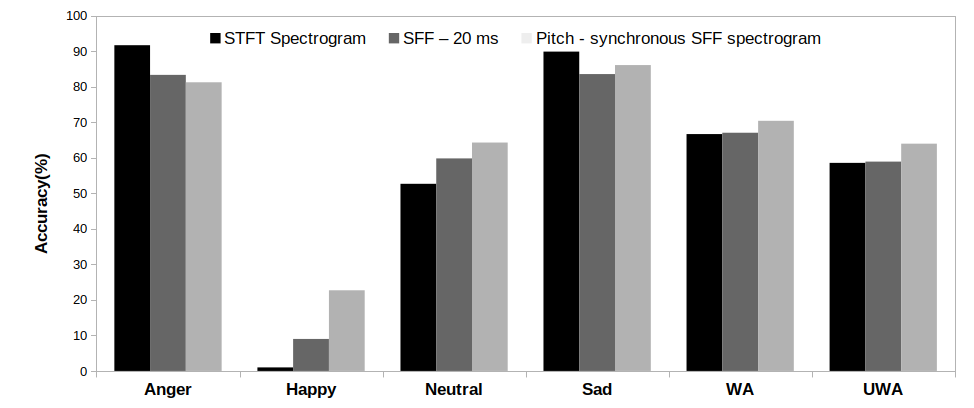}
\caption{Emotion classification performance(\%) using STFT spectrogram, SFF-20 ms spectrogram and pitch-synchronous SFF spectrogram }
\label{accuracyplot}
\end{figure*}

\begin{table}[b]
\centering
\caption{Comparison of Weighted accuracy and Unweighted accuracy of the proposed method with the baseline method}\label{T4}
 \begin{tabular}{ |c|c|c|c|c| }

 \hline
 & Representation & UWA  & WA \\
 \hline
 \multirow{2}{*}
{Proposed} & Pitch-Synchronous SFF Spectrogram & 63.95 & 70.4
\\\cline{2-4}

 & SFF-20 ms & 59.09 & 67.84\\
 
 \hline 
 Baseline & STFT-Spectrogram &  58.55 & 67.04 \\
 \hline
 Satt et al.\cite{satt2017efficient}& STFT Spectrogram & 56.6 & 66.1 \\
 \hline
\end{tabular}

\end{table}

Regularization techniques have been used to prevent overfitting. Regularization makes slight changes in the learning algorithms such that the model can generalize better, which improves the performance of the model on unseen data. This also helps in faster convergence. We have used multiple regularization techniques, namely batch normalization, dropout, LOSO cross validation, model selection, and early stopping. Batch normalization is used to normalize the convolution layer outputs to reduce the vanishing gradient problem in the activation layer. Batch normalization reduces the dependency of the designed network  on weight initialization. It improves the gradient flow through the network and also adds some amount of regularization into the network since the empirical means and variances are calculated using samples from mini-batches. Therefore, batch normalization helps to increase the accuracy during testing. 

Dropout has also been used as a regularization technique in this work. At each iteration, a random number of nodes are selected and removed. Thus, different number of nodes at every iteration result in different outputs. It is similar to an ensemble technique used in machine learning that captures more randomness. LOSO cross-validation is also used to avoid overfitting of the model. Experimental results show that the learned network  is better  generalized after cross-validation. The model was saved on the basis of maximum validation accuracy. The early stopping criteria is used to stop the model based on the value of the patience parameter. Patience is the number of epochs for which a model should continue training in spite of no improvement in the validation accuracy. The value of patience is chosen as five. In our experiments, several methods for reducing overfitting were used but overfitting was not completely avoided.

We have compared the proposed pitch-synchronous SFF spectrogram model with the following:
\begin{itemize}
	\item State-of-the-art result on STFT spectrogram using CNN as proposed by Satt, Rozenberg and Hooory \cite{satt2017efficient}.
	\item The batch normalized model of state-of-the-art result \cite{satt2017efficient} on the STFT spectrogram. We call this baseline model because any improvement of the proposed model over baseline is purely due to the pitch-synchronous SFF technique. 
	\item SFF-20 ms spectrogram. This represents the use of the SFF spectrogram with the traditional block processing technique.   
\end{itemize}

Table \ref{T4} shows the corresponding comparison results.
The used regularization techniques improved WA and UWA by $+1.95$\% and $+.94$\% respectively over state-of-the-art \cite{satt2017efficient}. There is further, albeit small, improvement when using the SFF-20ms spectrogram. The best results are obtained for the proposed pitch-synchronous SFF spectrogram with accuracy values of 63.95\% (UWA) and 70.4\% (WA). These correspond to improvements of 7.35\% (UWA) and 4.3\% (WA) over the state-of-the-art STFT spectrogram \cite{satt2017efficient}. Looking at the results for the proposed  pitch-synchronous SFF spectrogram and the SFF-20ms spectrogram, it can be said that a significant portion of improvement over state-of-the-art STFT spectrogram is due to logical windowing between two successive GCI locations, however, it must be noted that such logical windowing is not possible with the STFT spectrogram. This is because the STFT spectrogram does not obtain information at each instant of time.

The confusion matrices of the baseline STFT spectrogram and the pitch-synchronous SFF spectrogram are shown in Tables \ref{T2} and \ref{T3} respectively. The accuracy of anger and sad emotions are satisfactory using the baseline STFT spectrogram. But, it fails to categorize happy emotion.  All the samples of happy emotion are predicted as neutral using the STFT spectrogram. The same is also true for state-of-the-art results in \cite{satt2017efficient} (the corresponding confusion matrix has not been shown here). However, the pitch-synchronous SFF spectrogram categorizes $22.7$\% of happy emotions correctly. The accuracy of happy and neutral emotions are improved by $+22.7$\% and $+11.61$\% respectively in the case of the pitch-synchronous SFF spectrogram. This can be considered as a big plus for the the pitch-synchronous SFF spectrogram when compared to the STFT spectrogram.  The accuracies of sad and anger emotions are approximately same for the STFT  and pitch-synchronous SFF spectrograms. 

 \begin{table}[tp]
\centering
 \begin{center}
 \caption{Confusion matrix of STFT spectrogram in percentage}\label{T2}
 \begin{tabular}{ |c|c|c|c|c| } 
 \hline
 Emotion & Anger & Happy & Neutral & Sad \\
 \hline 
 Anger & \textbf{91.67} &0 & 0 & 8.33 \\
 \hline
 Happy & 40.91 & \textbf{0} & 50 & 9.1 \\
 \hline
 Neutral & 17.86 &3.57 & \textbf{52.67} & 25.89 \\
 \hline
 Sad & 0 & 0 & 10.12 & \textbf{89.87} \\
 \hline
 \end{tabular}
 \end{center}
 \end{table}

\begin{table}[tp]
\centering
\begin{center}
\caption{Confusion matrix for proposed pitch-synchronous SFF spectrogram in percentage}\label{T3}
\begin{tabular}{ |c|c|c|c|c| } 
 \hline
 Emotion & Anger & Happy & Neutral & Sad \\
 \hline 
 Anger & \textbf{81.25} &4.16 & 14.58 & 0 \\
 \hline
 Happy & 27.27 & \textbf{22.7} & 45.45 & 4.55 \\
 \hline
 Neutral & 6.25 & 9.82 & \textbf{64.28} & 19.64 \\
 \hline
 Sad & 0 & 0 & 13.92 & \textbf{86.07} \\
 \hline
 \end{tabular}
 \end{center}
 \end{table}

The class accuracies of each emotion for the STFT,  SFF-20 ms  and pitch-synchronous SFF spectrograms are shown in Fig. \ref{accuracyplot}. The SFF-20 ms spectrogram is computed from the  SFF envelope. The framing procedure is same as the traditional STFT spectrogram where 20 ms frame size with 50\% overlapping of previous frame is taken. The average of all the samples in a frame of an SFF envelope is computed. Using the SFF-20 ms spectrogram, an improvement of 2.49\% (UWA) and 1.74\% (WA)   was observed as compared to the state-of-the-art STFT spectrogram using CNN \cite{satt2017efficient}. The corresponding improvement for pitch-synchronous SFF spectrogram is 7.35\% (UWA) and 4.3\% (WA).



\section{Conclusion and Future Work}
This paper highlights the drawbacks of feature representation of the existing STFT spectrogram for SER. We have proposed a novel pitch-synchronous SFF spectrogram for SER that overcomes these drawbacks. We have attempted to solve the problem of SER using deep convolutional neural networks.
 Our proposed architecture consists of three CNN blocks followed by a fully connected layer and an output layer for detecting four emotions (i.e. Anger, Happy, Neutral, Sad).  On IEMOCAP dataset, the proposed pitch-synchronous SFF spectrogram achieved an improvement of $+7.35$\% and $+4.3$\% for weighted and unweighted accuracy values respectively over  state-of-the-art STFT spectrogram representation using CNN. Specially, the proposed pitch-synchronous SFF spectrogram recognized 22.7\% of the happy emotion samples correctly, whereas this number was 0\% for  state-of-the-art STFT spectrogram representation using CNN. 
 
 Pitch-synchronous SFF spectrogram can be used in various other applications such as speaker identification \cite{kekre2012speaker, wang2017spectral}, speaker verification \cite{yeh2002method}, audio classification \cite{zeng2019spectrogram} etc. The SFF output has  high SNR values of speech in the time-frequency domain. Our future work is to explore these properties of SFF to develop a robust speech emotion recognition against degradations due to noise.
 
\section*{ACKNOWLEDGEMENTS}
Akshay Deepak has been awarded Young Faculty Research
Fellowship (YFRF) of Visvesvaraya PhD Programme of
Ministry of Electronics \& Information Technology, MeitY,
Government of India. In this regard, he would like to
acknowledge that this publication is an outcome of the R\&D
work undertaken in the project under the Visvesvaraya PhD
Scheme of Ministry of Electronics \& Information Technol-
ogy, Government of India, being implemented by Digital
India Corporation (formerly Media Lab Asia).

\bibliographystyle{plain}
\bibliography{acnref}
\end{document}